\documentclass[preprint,12pt]{elsarticle}

\usepackage{amssymb}
\usepackage{amsmath}
\usepackage{graphicx} 
\usepackage{amsfonts}
\usepackage{natbib}
\usepackage{textcomp}
\usepackage{float}
\usepackage{subcaption}
\usepackage{textcomp}
\usepackage{acronym}
\usepackage{pdflscape}
\usepackage{comment}
\usepackage{array}
\usepackage{booktabs}
\usepackage[hidelinks]{hyperref}
\newcolumntype{P}[1]{>{\centering\arraybackslash}p{#1}}
\newcolumntype{M}[1]{>{\centering\arraybackslash}m{#1}}
\usepackage{etoolbox}
\usepackage{cleveref}
\usepackage{eurosym}
\usepackage{indentfirst}
\usepackage{adjustbox}
\usepackage{nicefrac}
\usepackage{multirow}
\usepackage{subcaption}
\usepackage{bbding}
\usepackage[table]{xcolor}
\usepackage{pifont}

\RequirePackage[%
  acronym,
  automake,
  nogroupskip,
  nopostdot,
  nonumberlist,
  toc,
  ]{glossaries}

\newglossarystyle{mylong}{%
  \setglossarystyle{long}
    {\begin{supertabular}{p{1.7cm}p{\dimexpr\linewidth\relax}}}%
    {\end{supertabular}}%
}

\makeglossaries

\newacronym{uc}{UC}{unit commitment}
\newacronym{milp}{MILP}{mixed-integer linear programming}
\newacronym{rocof}{RoCoF}{rate of change of frequency}
\newacronym{ips}{IPS}{island power systems}
\newacronym{scuc}{SCUC}{security-constrained unit commitment}
\newacronym{fcuc}{FCUC}{frequency-constrained unit commitment}
\newacronym{pfcuc}{P-FCUC}{preventive frequency-constrained unit commitment}
\newacronym{cfcuc}{C-FCUC}{corrective frequency-constrained unit commitment}
\newacronym{minlp}{MINLP}{mixed-integer non-linear programming}
\newacronym{rc}{RC}{reserve constrained}
\newacronym{buc}{BUC}{base case unit commitment}
\newacronym{sfr}{SFR}{system frequency response}
\newacronym{ufls}{UFLS}{under-frequency load shedding}
\newacronym{kpi}{KPI}{key performance indicator}
\newacronym{res}{RES}{renewable energy sources}
\newacronym{pfr}{PFR}{primary frequency response}

\usepackage[intoc]{nomencl}
\usepackage{xstring}
\usepackage{xpatch}
\usepackage{eurosym}

\patchcmd{\thenomenclature}
  {\leftmargin\labelwidth}
  {\leftmargin\labelwidth\itemindent 1em }
  {}{}
\newcommand{\nomenclheader}[1]{%
  \item[\hspace*{-\itemindent}\normalfont\bfseries#1]}
\renewcommand\nomgroup[1]{%
  \IfStrEqCase{#1}{%
   {A}{\nomenclheader{Indices and sets}}
   {B}{\nomenclheader{Parameters}}
   {C}{\nomenclheader{Variables}}
  }%
}

\makenomenclature

\nomenclature[A]{\(\ell\)}{Index of the lost unit}
\nomenclature[A]{\(\mathcal{T}\)}{Set of all time intervals}
\nomenclature[A]{\(i\)}{Index of generators}
\nomenclature[A]{\(t\)}{Index of time intervals}
\nomenclature[A]{\(s\)}{Alias index for time intervals}
\nomenclature[A]{\(j\)}{Index of features}
\nomenclature[A]{\(r\)}{Index of piecewise linearization segments}
\nomenclature[A]{\(\mathcal{R}\)}{Set of all piecewise linearization segments}
\nomenclature[A]{\(\mathcal{L}\)}{Set of all resulting leaves}
\nomenclature[A]{\(J\)}{Set of all features}

\nomenclature[B]{\(\text{UT}\)}{Minimum up-time of generators [hours]}
\nomenclature[B]{\(\text{DT}\)}{Minimum downtime of generators [hours]}
\nomenclature[B]{\(P^{\text{max}}_i\)}{Maximum power output of unit $i$ [MW]}
\nomenclature[B]{\(P^{\text{min}}_i\)}{Minimum power output of unit $i$ [MW]}
\nomenclature[B]{\(R^{\text{up}}_i\)}{Maximum ramp-up of unit $i$ [MW]}
\nomenclature[B]{\(R^{\text{down}}_i\)}{Maximum ramp-down of unit $i$ [MW]}
\nomenclature[B]{\(g^{w}_t\)}{Wind generation at time $t$ [MW]}
\nomenclature[B]{\(g^{s}_t\)}{Solar generation at time $t$ [MW]}
\nomenclature[B]{\(\mathcal{D}_t\)}{Demand at time $t$ [MW]}
\nomenclature[B]{\(f_0\)}{Nominal frequency [Hz]}
\nomenclature[B]{\(T\)}{Delivery time of units [s]}
\nomenclature[B]{\(K_i\)}{Turbine-governor gain of unit $i$}
\nomenclature[B]{\({\hat K}_i\)}{Turbine-governor gain of unit $i$ divided by its $T$}
\nomenclature[B]{\(M\)}{A sufficiently big positive number}
\nomenclature[B]{\(D\)}{Load damping factor}
\nomenclature[B]{\(\Delta f_{\text{crit}}^{'}\)}{Critical rate of change of frequency}
\nomenclature[B]{\(\Delta f_{\text{crit}}^{\text{ss}}\)}{Critical steady-state frequency [Hz]}
\nomenclature[B]{\(\Delta f_{\text{crit}}^{\text{nadir}}\)}{Critical frequency nadir [Hz]}
\nomenclature[B]{\(\theta_j\)}{Coefficient of feature $j$}
\nomenclature[B]{\(\theta_0\)}{Intercept of the logistic function}
\nomenclature[B]{\(\alpha_j\)}{Linear regression coefficients}
\nomenclature[B]{\(\beta_j\)}{Logistic regression coefficients}
\nomenclature[C]{\(p^{\text{\tiny UFLS}}_{\ell,t}\)}{Amount of UFLS after outage of unit $\ell$ at time $t$ [MW]}
\nomenclature[C]{\(v\)}{Start-up variable [$\in$\{0,1\}]}
\nomenclature[C]{\(w\)}{Shut-down variable [$\in$\{0,1\}]}
\nomenclature[C]{\(p\)}{Power variable [MW]}
\nomenclature[C]{\(c^{\text{g}}\)}{Generation costs [\euro]}
\nomenclature[C]{\(c^{\text{suc}}\)}{Start-up costs [\euro]}
\nomenclature[C]{\(u\)}{Commitment variable [$\in$\{0,1\}]}
\nomenclature[C]{\(r_{i,t}\)}{Reserve provided by unit $i$ at time $t$ [MW]}
\nomenclature[C]{\(H_{\ell}\)}{Weighted summation of inertia constants of online units after outage of unit $\ell$ [s.MW]}
\nomenclature[C]{\(p_{\ell}\)}{Power lost after outage of unit $\ell$ [MW]}
\nomenclature[C]{\(P_m\)}{Mechanical power [MW]}
\nomenclature[C]{\(P_e\)}{Electrical power [MW]}
\nomenclature[C]{\(R_{\ell}\)}{Sum of available reserve after outage of unit $\ell$ [MW]}
\nomenclature[C]{\(\tau\)}{Time variable}
\nomenclature[C]{\(x\)}{Vector of features}
\nomenclature[C]{\(p^{\text{\tiny crit}}_{\ell,t}\)}{Maximum power outage with no UFLS [MW]}
\nomenclature[C]{\(z_{\ell e}\)}{Leaf membership variable [$\in$\{0,1\}]}

\journal{Electric Power Systems Research}

\begin{document}

\begin{frontmatter}



\title{A Comparative Study on Frequency-Constrained Unit Commitment Approaches in Island Power Systems}

\author[inst1]{Miad Sarvarizadeh} 
\author[inst1]{Mohammad Rajabdorri}
\author[inst1]{Enrique Lobato}
\author[inst1]{Lukas Sigrist}

\affiliation[inst1]{organization={IIT, Comillas Pontifical University},
            city={Madrid},
            country={Spain}}

\begin{abstract}
The increasing penetration of renewable energy sources reduces rotating inertia and even frequency control capacity, affecting frequency stability. This challenge is significant in \gls{ips} that already suffer from low inertia and frequency control capacity.
This paper presents a comparative study on different \gls{fcuc} formulations applied to \gls{ips}. Then, by considering under-frequency load shedding as a significant measure of frequency stability in \gls{ips}, two indices are presented to fully compare the formulations from system benefits and computational burden perspectives. Simulations conducted on a real Spanish island show that the data-driven corrective \gls{fcuc} formulation has the most advantages among other formulations.
\end{abstract}
\glsresetall


\begin{highlights}
\item A comparison framework is provided to compare four frequency-constrained unit commitment methods with a base case unit commitment formulation, regarding the system benefits and computational burden perspectives.
\item Two performance indices are introduced to help analyze and compare the formulations more effectively by considering the under-frequency load shedding reduction.
\item Each formulation is implemented on a real Spanish island power system illustrating its effectiveness in real-world systems and helping operators choose the suitable method in different circumstances.
\end{highlights}


\end{frontmatter}

\printnomenclature

\section{Introduction}\label{intro}

\subsection{Motivation}\label{background}
The \gls{uc} problem involves determining the most economically efficient scheduling of generation units in a power system. The goal of the \gls{uc} problem is to minimize total system operation costs while considering system operation constraints, and all technological limitations of generation units. However, traditional \gls{uc} formulations tend to ignore frequency-related constraints. Frequency stability can be added to the \gls{uc} problem leading to \gls{fcuc} formulations.

The small size of \gls{ips} indicates that fewer generating units are available, resulting in lower inertia and frequency control capacity. Furthermore, the spinning reserve requirements are more demanding concerning the size of the units, leading to higher overall operation costs. Increasing the use of \gls{res} in \gls{ips} could lower operational costs and improve sustainability. However, with more \gls{res} the already limited inertia of the system will be lower, and the frequency response following a generator outage will have less favorable characteristics \cite{sigrist2016island}. Consequently, preventing \gls{ufls} entirely by setting conservative frequency nadir thresholds would not be feasible in small \gls{ips}. Therefore, \gls{ufls} is inevitable for large outages in \gls{ips} \cite{rajabdorri2022robust}.

Including frequency-related constraints into the \gls{uc} problem of \gls{ips} causes changes from two different perspectives. From the operational perspective, the operation costs will be higher as a trade-off for better post-contingency frequency performance, improving frequency stability. The second perspective is the computational burden. Frequency constraints often add a lot of new variables and constraints to the problem and could result in higher solution times compared to the traditional \gls{uc} formulations.

Different indices have been considered to evaluate the performance of an \gls{fcuc} formulation for \gls{ips}. The amount of \gls{ufls} is a significant measure of frequency stability since the \gls{ufls} scheme acts when frequency stability is put at risk, i.e., when frequency deviations or \gls{rocof} exceed certain thresholds by the system operator. Accordingly, the ultimate goal of the \gls{fcuc} formulations in \gls{ips} is to minimize the total possible amount of \gls{ufls} after outages while imposing minimum extra costs and computational burden on the problem. This concept is the foundation of the comparison framework presented in this paper.

\subsection{Literature review}\label{review}
 To ensure power grid frequency stability following disturbances, three primary frequency indicators (\gls{rocof}, frequency nadir, and quasi-steady-state frequency) are constrained to thresholds \cite{mancarella2021fragile}. Adding these constraints limits the characteristics of the post-contingency frequency performance, thus forming \gls{pfcuc} formulations. The frequency-related constraints are modeled utilizing analytical or data-driven methods.
 
 Several papers have presented analytical \gls{pfcuc} formulations. Bernstein polynomials are used in \cite{zhou2023frequency} to approximate the differential-algebraic equations describing the power system. Before being implemented in the \gls{uc}, the differential equations are first transformed into a set of linear equations. The dynamic response solution offers suitable restrictions by utilizing the characteristics of Bernstein polynomials.
 \cite{tuo2022security} presents a location-based \gls{uc} formulation that is \gls{rocof}-constrained and focuses on location-based frequency characteristic differentiation. Furthermore, the cost-benefit impact of virtual inertia is investigated.
 Analytical constraints for \gls{rocof} and frequency nadir are derived in \cite{wu2024unit} based on the center of inertia concept. Moreover, random sampling of generation unit combinations is utilized to obtain lower bounds for the frequency
constraints. 
An analytical method for modeling frequency stability constraints that includes the frequency response from \gls{res} is presented in \cite{zhang2023frequency}. The suggested \gls{minlp} problem is solved using decomposition techniques.
In \cite{ferrandon2022inclusion}, a separable programming technique is used to model the nonlinear frequency nadir constraint in the \gls{uc} problem. The frequency nadir constraint is approximated using a piece-wise linear function.

Some studies have utilized data-driven approaches to formulate \gls{pfcuc} problems. 
\cite{liu2024frequency} presents a frequency nadir constraint using linear regression. Requirements for minimum inertia or frequency nadir are included in this model. To validate the presented model the Maui Island system is used.
An extreme learning machine model is presented in \cite{liu2023modeling} to obtain a linearized frequency nadir constraint. Furthermore, in the day-ahead \gls{uc} problem, the droop gains and reserves of available generators are optimized to consider the variations in frequency insecurity.
A deep neural network-based data-driven framework for predicting frequency nadir is described in \cite{zhang2021encoding} and incorporated into \gls{milp} formulations. To avoid \gls{ufls}, a region-of-interest active sampling is used to choose samples with frequency nadirs near the threshold.
In \cite{sang2023conservative}, a sparse neural network technique based on a conservative sparse neural network is presented for approximating step-wise constraints and frequency nadir. An \gls{milp} representation of these constraints is obtained to add them to the \gls{uc} problem. A data-driven approach in modeling non-linear frequency nadir constraint in the \gls{uc} problem using logistic regression and support vector machine is presented in \cite{rajabdorri2023inclusion}. The results are provided for a real \gls{ips} and compared with an analytical method.

Corrective measures such as \gls{ufls} are utilized in power systems as a last resort to alter frequency decay \cite{zou2020}. A \gls{ufls} scheme automatically sheds load based on system measurements. According to \cite{sigrist2018}, these automatic \gls{ufls} schemes can be categorized as conventional or advanced. Plenty of studies have focused on designing \gls{ufls} schemes. However, few papers have studied \gls{ufls} as a corrective measure in a \gls{fcuc} formulation.

Authors in \cite{ceja2012under} examined the optimal \gls{ufls} schemes to reduce the amount of shed load using predetermined generating commitment states. Similarly, \cite{bhana2015commitment} explores the optimal commitment of interruptible loads to ensure adequate primary frequency response, but does not consider the commitment of generating units.
One of the initial studies on incorporating \gls{ufls} in the \gls{fcuc} formulations is proposed in \cite{teng2017full}. An analytical estimation of \gls{ufls} is calculated considering linear reserve provision of the units in response to a known outage. \gls{ufls} is divided into blocks, with one nadir constraint per block. Then binary variables are utilized to determine the optimal \gls{ufls} level for scheduling.
A framework to co-optimize UFLS and fast and slow frequency responses is provided in \cite{o2021probabilistic}. The resulting non-convex estimation of \gls{ufls} is approximated with a second-order cone, which is fairly accurate. In \cite{rajabdorri2024unit} an analytical method to include \gls{ufls} that gives a \gls{cfcuc} formulation is presented. Similarly, it is assumed that the units provide reserve linearly and the amount of power outage is known.

Estimating and co-optimizing \gls{ufls} using data-driven approaches is proposed in \cite{RAJABDORRI2025data}. A learning process to estimate the amount of \gls{ufls} is presented, considering conventional schemes due to their extensive utilization.
This formulation also considers the maximum \gls{rocof} constraint, so it can be viewed as a hybrid formulation that proposes both corrective and preventive measures.
This formulation is studied on a real-world Spanish \gls{ips} demonstrating its effectiveness in estimating real amounts of \gls{ufls}.

As indicated, many studies have provided \gls{fcuc} formulations and successfully improved the post-contingency frequency performance of \gls{ips}. These studies can be categorized into analytical or data-driven, preventive or corrective \gls{fcuc}. However, most of these studies are isolated studies that ignore comparisons to other methods and overlook the computational burden they impose on current formulations.
This study tries to address the practical limitations of \gls{fcuc} formulations by comparing four different methods (one for each category) from two different perspectives: system benefits and computational burden.

\begin{figure}[t!]
    \centering
    \includegraphics[width=1\linewidth]{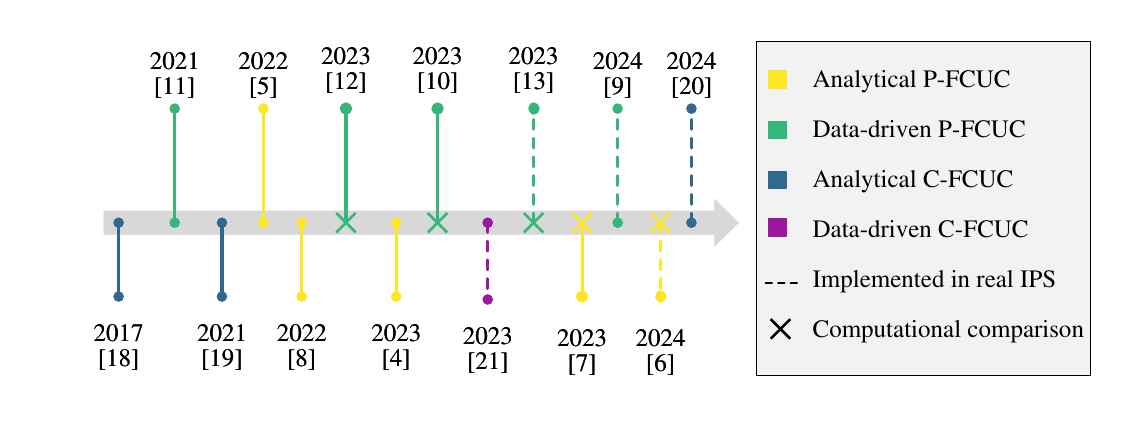}
    \caption{Timeline of the reviewed literature}
    \label{fig:timeline}
\end{figure}

\Cref{fig:timeline} illustrates a timeline of the reviewed literature. This figure highlights the methods used in each study and shows whether they are applied to real-world \gls{ips}. Additionally, \Cref{fig:timeline} shows if any analysis regarding the computational burden of the formulation is presented in each paper.  It is important to note that most studies that provide analysis of the computational burden, fall short of analyzing it according to the benefits they provide to the system and compare them to other methods possible.

This paper provides a framework to compare four different \gls{fcuc} methods implemented in small \gls{ips}, with a \gls{buc} formulation. Comparing \gls{fcuc} methods is crucial because each one employs different simplifying assumptions and varies in computational complexity. It is essential to evaluate how accurately these methods approximate more detailed dynamic simulations, which can provide a more realistic representation of system behavior. The method in \cite{ferrandon2022inclusion} is used as a representative of analytical \gls{pfcuc}, whereas the one in \cite{rajabdorri2023inclusion} serves as a representative of data-driven \gls{pfcuc}. An analytical \gls{cfcuc} formulation is represented by \cite{rajabdorri2024unit}. We adopted the methodology from \cite{RAJABDORRI2025data}, which estimates the amount of \gls{ufls} using regression trees, and integrated it into the \gls{uc} problem. This implementation serves as a data-driven \gls{cfcuc} in our comparison. The code used for solving all of these formulations is available on \url{https://github.com/Miadsrv/FrequencyConstrainedUC}.
It is important to note that although these formulations do not represent every possible formulation of \gls{fcuc}, they are the main ones that are proposed for \gls{ips} considering the unique characteristics of these systems. Therefore, this study provides clear insights into the scheduling problem in \gls{ips}, considering frequency stability.
\subsection{Contributions and structure}\label{contribution}
The main contributions of this paper are the following:
\begin{itemize}
    \item A comparison framework is provided to compare four different \gls{fcuc} methods with a base case \gls{uc} formulation for \gls{ips}, regarding the system benefits and computational burden perspectives.
    \item The methodology in \cite{RAJABDORRI2025data} estimates the amount of \gls{ufls} triggered by real-world schemes. Here, the estimation is incorporated into the \gls{uc} problem, forming a \gls{cfcuc} model.
    \item Two performance indices are introduced to help analyze and compare the formulations more effectively from the system benefits and computational burden viewpoints by considering the amount of \gls{ufls} reduction.
    \item Each formulation is implemented on a real Spanish \gls{ips} illustrating its effectiveness in real-world \gls{ips} to help operators choose the suitable method in different circumstances.
\end{itemize}

The remainder of this paper is organized as follows: first of all, the mathematical formulation of base case \gls{uc} problem and four different \gls{fcuc} methods are provided in \Cref{math}. \Cref{benchmark} describes the comparison framework and the \glspl{kpi}. Then, \Cref{results} provides the obtained results. Finally, \Cref{conclusions} gives some conclusions.

\section{Mathematical formulation of models}\label{math}
The \gls{uc} problem is formulated to minimize the operation cost of generators and considers various constraints. This section presents the complete formulation of \gls{buc} as the benchmark formulation. Then, the four selected \gls{fcuc} formulations are presented, each representing a different category of formulations.

\subsection{Base case formulation}\label{base}
The \gls{buc} model is a typical representation of the \gls{uc} problem that includes \cref{of,bin1,bin2,ut,dt,pmin,pmax,ru,rd,pb,res}.
The objective function presented in \cref{of} minimizes the total operation cost of the system including generation and start-up costs.

\begin{subequations}
\begin{align}
    \min_{u,p}\ c^{\text{g}}(p)&+c^{\text{suc}}(u)\label{of}
    \\
    u_{i,t}-u_{i,t-1}&=v_{i,t}-w_{i,t} &&\text{\footnotesize $\forall i,\; \forall t$}\label{bin1} 
    \\
    v_{i,t}+w_{i,t}&\leq1 &&\text{\footnotesize $\forall i,\;\forall t$}\label{bin2} 
    \\
    \sum_{s=t-\text{UT}_i+1}^{t}v_{i,s}&\leq u_{i,t} &&\text{\footnotesize $t\in\{\text{UT}_i,\dots, \mathcal{T}\}$}\label{ut} 
    \\
   \sum_{s=t-\text{DT}_i+1}^{t}w_{i,s}&\leq 1-u_{i,t} &&\text{\footnotesize $t\in\{\text{DT}_i,\dots, \mathcal{T}\}$}\label{dt} 
   \\
   p_{i,t}&\geq P^{\text{min}}_i u_{i,t} &&\text{\footnotesize $\forall i,\; \forall t$}\label{pmin}  
   \\
    p_{i,t}+r_{i,t}&\leq P^{\text{max}}_i u_{i,t} &&\text{\footnotesize $\forall i,\; \forall t$}\label{pmax} 
    \\
    p_{i,t}-p_{i,t-1}&\leq R^{\text{\tiny{up}}}_i &&\text{\footnotesize $\forall i,\; \forall t$}\label{ru} 
    \\
   p_{i,t-1}-p_{i,t}&\leq R^{\text{\tiny{down}}}_i &&\text{\footnotesize $\forall i,\; \forall t$}\label{rd} 
   \\
   \bigg(\sum_{i}p_{i,t}\bigg) +g^w_t+g^s_t& = \mathcal{D}_t &&\text{\footnotesize $\forall t$}\label{pb}
   \\
   \sum\limits_{i\neq \ell}r_{i,t}&\geq p_{\ell,t}&&\text{\footnotesize $\forall t,\; \ell$}\label{res}
\end{align}
\end{subequations}
\Cref{bin1,bin2} represent the binary logic of the \gls{uc} problem. The minimum up-time and down-time constraints are presented in \cref{ut,dt}. Each generator's output and reserve are constrained by the minimum and maximum capacity and ramp rate bounds through \cref{pmin,pmax,ru,rd}. \Cref{pb} is the power balance constraint. The reserve constraint is presented in \cref{res} that makes sure there is enough reserve to compensate for the active power disturbance of generating unit
$\ell$.

\subsection{Preventive FCUC formulations}\label{preventive}
The dynamics of the generator rotor in power systems are usually described by the swing equation of \cref{swing}.
\begin{align}\label{swing}
\frac{2H}{f_0}\frac{d\Delta f(t)}{dt}&+D\mathcal{D}_t\Delta f(t)=P_m-P_e  
\end{align}

Formulations of \gls{pfcuc} aim to incorporate the system dynamics by analyzing the \cref{swing} equation and reflecting these dynamics within the \gls{uc} problem. The objective of these formulations is to schedule generation to mitigate the risk of poor frequency responses following $N-1$ contingencies, thereby enhancing system reliability. Preventive \gls{fcuc} formulations incorporate constraints to ensure that key measures (\gls{rocof}, quasi-steady-state frequency, and frequency nadir) remain within acceptable limits.

The maximum \gls{rocof} constraint is presented in \cref{rocof} indicates the minimum level of inertia requirement based on a maximum \gls{rocof} allowed. This linear constraint can be added directly to the \gls{buc} formulations.
\begin{align}\label{rocof}
   {H}_\ell&\geq \frac{p_\ell f_0}{2\Delta f_{\text{crit}}}&&\text{\footnotesize $\forall t,\; \ell$}  
\end{align}

For the quasi-steady-state frequency following an outage, it is assumed that it has stabilized without the secondary frequency control activation. To ensure that the quasi-steady-state frequency remains within acceptable limits, \cref{qss} can be derived from the swing equation. This equation is also linear and is added directly to the formulations.
\begin{align}\label{qss}
  \sum\limits_{i\neq \ell}r_{i,t}&\geq P_\ell-D\mathcal{D}_t\Delta f^{\text{ss}}_{\text{crit}}&&\text{\footnotesize $\forall t,\; \ell$}  
\end{align}

The frequency nadir constraint is more complicated to add to the \gls{uc} problem. Analytical and data-driven methods are the two main approaches to modeling the frequency nadir which are discussed in the following.

\subsubsection{Analytical nadir constraints}\label{anadir}
The analytical model of frequency nadir discussed here is based on \cite{ferrandon2022inclusion} where separable programming is utilized. A key assumption in this method, widely used in the literature, is that each unit's reserve increases linearly and achieves its peak output within $T$ seconds as indicated in \cref{reslin}. The frequency nadir is assumed to happen before $T$.
\begin{align}\label{reslin}
r_{i,t}(\tau) &=
    \begin{cases}
      \frac{r_{i,t}\tau}{T} & \text{if $\tau\leq T$}\\
      r_{i,t} & \text{if $\tau > T$}
    \end{cases}       
\end{align}

The resulting non-convex frequency nadir constraint is presented in \cref{nadir}.
Change of variables and separable programming are utilized to linearize this constraint. 
\begin{align}\label{nadir}
    H_\ell R_\ell - \frac{f_0Tp_\ell^2}{4\Delta f^{nadir}_{crit}}+\frac{D\mathcal{D}_tTp_\ell f_0}{4}\geq0&&\text{\footnotesize $\forall t,\; \ell$}
\end{align}

The product $H_\ell R_\ell$ is rewritten using the following variable change.
\begin{subequations}\label{varc}
\begin{align}
    H_\ell\alpha R_\ell\beta=x_1^2-x_2^2 \label{varchange}
    \\
    \frac{x_1+x_2}{\alpha}=H_\ell
    \\
    \frac{x_1-x_2}{\beta}=R_\ell
\end{align}
\end{subequations}

Then, by substituting \cref{varchange} in\cref{nadir}, the resulting frequency nadir equation can be written as the following:
\begin{align}\label{modnadir}
    \frac{x_1^2-x_2^2}{\alpha\beta} - \frac{f_0Tp_\ell^2}{4\Delta f^{\text{nadir}}_{\text{crit}}}+\frac{D\mathcal{D}_tTp_\ell f_0}{4}\geq0&&\text{\footnotesize $\forall t,\; \ell$}
\end{align}

The remaining non-linear terms are $p_\ell^2$, $x_1^2$, and $x_2^2$, which can be linearized quite straightforwardly. The linearization process of these terms is presented in detail in \cite{ferrandon2021secure}.

\subsubsection{Data-driven nadir constraints}\label{mlnadir}
Another approach to include frequency nadir constraint into the \gls{uc} problem is based on the data-driven estimation of the frequency nadir by using machine learning techniques. The data-driven method discussed here is described in detail in \cite{rajabdorri2023inclusion}. This approach includes several steps including data generation, labeling the data set with the corresponding nadir frequency, and training the machine learning model. Here a summary of the process is presented.

A well-structured dataset of features and labels is essential to predict the frequency nadir in \gls{uc} problems. The features represent operating points and are paired with labels representing the frequency nadir measurements after all possible outages. These labels are derived using dynamic \gls{sfr} models. Labels can be numeric (e.g. frequency in Hz) or categorical (e.g. tolerable or not).

The selected features must efficiently predict the label without overcomplicating the model. Only the most relevant features accessible in the \gls{uc} problem must be chosen. The selected features for predicting frequency nadir adequacy are the sum of available inertia ($H_\ell$), the weighted gain of the turbine-governor model ($\hat K^{s}_{\ell,t}$) defined in \cref{equ_k}, lost power ($p_\ell$), and the sum of available reserve ($R_\ell$) after the outage of generator $\ell$. Then the \gls{sfr} model is used to label the data set with the corresponding frequency nadir of each possible outage. The \gls{sfr} model utilizes a second-order approximation to represent the turbine-governor system of each generating unit. Details of the \gls{sfr} model can be found in \cite{sigrist2016island}. Note that any other power system model can be used instead of the \gls{sfr} model.
\begin{equation}\label{equ_k}
    \hat K^{s}_{\ell,t} = \sum\limits_{i \ne \ell}{\frac{K_i}{T_i} u_{i,t}}
\end{equation}

Here, a logistic regression model is utilized. The labels classify data points, where $+1$ represents acceptable outcomes, which means that the frequency nadir is in the acceptable range. Alternatively, $0$ represents the incidents for which their corresponding frequency nadir is unacceptable. The machine learning model aims to learn a decision function $f_{\theta}(x)$ presented in \cref{eqmlnadir}, which produces positive values for acceptable points and negative values for unacceptable ones while minimizing misclassifications.
\begin{equation}\label{eqmlnadir}
    f_{\theta}(x)=\theta_0+\sum_{j=1}^J \theta_j x_j
\end{equation}
The selected features are replaced in \cref{eqmlnadir}, and \cref{eqmlnadir2} is obtained.
\begin{equation}\label{eqmlnadir2}
    \theta_0+\theta_1H_\ell+\theta_2\hat K^{s}_{\ell,t}+\theta_3p_\ell+\theta_4R_\ell\geq 0
\end{equation}
After calculating the parameters of the \cref{eqmlnadir2}, the learned decision function can be directly added to the \gls{milp} formulation as a constraint.

\subsection{Corrective FCUC formulations}\label{corrective}
The formulations covered up to this point are based on preventive measures ensuring the frequency remains within acceptable limits without taking corrective action. 
Depending on the system, the \gls{pfcuc} formulations might successfully prevent \gls{ufls} from happening according to the imposed critical frequency limits. This is achievable for sufficiently large systems but may lead to increased system operation costs in gls{ips}. In addition, for small \gls{ips}, \gls{ufls} occurrences are inevitable. The smaller size implies that fewer generating units are connected, indicating that the inertia of the system and the overall frequency control capacity are lower.

Given these considerations, modeling \gls{ufls} inside the \gls{uc} problem is an effective approach that enables the utilization of inevitable \gls{ufls} occurrences along with spinning reserves. This would likely reduce operation costs as shown in \Cref{results} considering that a lower reserve could be scheduled. Also, the cost associated with activating \gls{ufls} can be minimized alongside the other operation costs in the objective function. Similar to the previous section, \gls{ufls} can be modeled and added to the \gls{uc} problem using analytical and data-driven approaches forming a \gls{cfcuc} formulation. 

\subsubsection{Analytical UFLS constraints}\label{aufls}
This formulation is based on \cite{rajabdorri2024unit}, estimating the amount of \gls{ufls} achievable by advanced \gls{ufls} schemes. First, the amount of generation loss that can lead to a frequency nadir that exceeds the preset acceptable threshold is presented as $p^{\text{crit}}_{\ell,t}$.
\begin{equation} \label{eqcrit}
    2H_{\ell,t}  \hat K^{s}_{\ell,t} (\Delta F^{\text{nadir}})^2 (S^{\text{base}})^2 =  (p^{\text{crit}}_{t,\ell})^2
\end{equation}
In \cref{eqcrit} $H_{\ell,t}$ is the some of available inertia after the outage of unit $\ell$, $\Delta F^{\text{nadir}}$ is the frequency nadir threshold, and $\hat K^{s}_{\ell,t}$ is defined in \cref{equ_k}. \Cref{eqcrit} contains two nonlinear terms.
$(p^{\text{crit}}_{t,\ell})^2$ can be linearized similar to the process presented in \cite{ferrandon2021secure}. Additionally, to linearize the binary to binary products resulting from $H_{\ell,t}  \hat K^{s}_{\ell,t}$ term, first, it can be written as \cref{linearbinary}.
\begin{equation}\label{linearbinary}
    H^{s}_{t,\ell} \times \hat K^{s}_{t,\ell} = \\
    \begin{bmatrix}
        \frac{K_1}{T_1} \\ \vdots \\ \frac{K_\ell}{T_\ell} \\ \vdots \\ \frac{K_n}{T_n}
    \end{bmatrix}^{-1}
    \begin{bmatrix}
    u_1u_1 & u_1u_2 & \dots & u_1u_\ell & \dots & u_1u_n\\
    \vdots & \vdots &  \vdots & \vdots  & \vdots  & \vdots\\
    u_\ell u_1 & u_\ell u_2 & \dots & u_\ell u_\ell & \dots & u_\ell u_n\\
    \vdots & \vdots &  \vdots & \vdots  & \vdots  & \vdots\\
    u_n u_1 & u_n u_2 & \dots & u_n u_\ell & \dots & u_n u_n\\
    \end{bmatrix}
    \begin{bmatrix}
        H_1 \\ \vdots \\ H_\ell \\ \vdots \\ H_n
    \end{bmatrix}
\end{equation}

The matrix containing binary-to-binary products is symmetric and the diagonal elements like $u_i u_i$ do not need to be linearized. The rest are linearized using the constraints presented in \cite{rajabdorri2024unit}.

If the outage is larger than $p^{\text{crit}}_{\ell,t}$, \gls{ufls} will be activated. Alternatively, if the outage is smaller than $p^{\text{crit}}_{\ell,t}$, no \gls{ufls} will be activated. The amount of \gls{ufls} is calculated in \cref{eqpufls}. This value of \gls{ufls} can be used as an estimation inside the \gls{uc} problem.
\begin{equation} \label{eqpufls}
    p^{\text{\tiny UFLS}}_{\ell,t} =
    \begin{cases}
        {p_{\ell,t} -  p^{\text{crit}}_{\ell,t}} & \text{if\;\;} p_{\ell,t} > p^{\text{crit}}_{\ell,t} \\
        0&\text{otherwise.}
    \end{cases}
\end{equation}

\Cref{eqpufls} is linearized here using the method presented in \cite{fischetti2018deep}. The following set of constraints substitute \cref{eqpufls}.
\begin{subequations} \label{equrw}
    \begin{align}
        p_{\ell,t} -  p^{\text{crit}}_{\ell,t}&=p^{\text{\tiny UFLS}}_{\ell,t}-e_{\ell,t}\\
        p^{\text{\tiny UFLS}}_{\ell,t}&\leq Mu'_{\ell,t}\\
        e_{\ell,t}&\leq M(1-u'_{\ell,t})
    \end{align}
\end{subequations}

In \cref{equrw}, $e_{\ell,t}$ is an auxiliary variable, $u'_{\ell,t}$ is a binary activation variable, and $M$ is a sufficiently large number. This set of constraints ensure that either $p^{\text{\tiny UFLS}}_{\ell,t}$ or $e_{\ell,t}$ is zero. This can alternatively be done using the indicator constraints that modern solvers widely accept.

Then, another constraint is applied to guarantee that each generating unit can deliver the reserve on time. \Cref{eqspeed} accounts for the distribution of the available reserve. It ensures that the remaining units after the outage will have enough headroom, relative to their speed of response. 
\begin{equation} \label{eqspeed}
    {p_{i,t} + \frac{{\hat K}_i u_{i,t}}{\hat K^{s}_{\ell,t}} p^{\text{crit}}_{\ell,t} \le P^{\text{max}}_i u_{i,t} \;\;\;\text{\footnotesize $\forall t,\; \ell$}}
\end{equation}

\Cref{eqspeed} contains several non-linear terms. The primary focus is on enforcing \cref{eqspeed} for online units. In this case, we assume that $u_{i,t}$ is equal to one and introduce an additional term to consistently satisfy the constraint for offline units (where $u_{i,t}$ equals zero). Accordingly, \cref{eqspeed} is updated as follows:
\begin{equation} \label{eqspeedmodified}
    {p_{i,t}\hat K^{s}_{\ell,t} + \hat{K}_i p^{c}_{\ell,t} \le P^{\text{max}}_i \hat K^{s}_{\ell,t} + M(1-u_{i,t}) \;\;\;\text{\footnotesize $\forall i \ne \ell$}}
\end{equation}

Here, $M$ represents a sufficiently large number, leaving the only remaining non-linear term $p_{i,t}\hat K^{s}_{\ell,t}$. Each resulting non-linear term is a binary in a continuous variable product as indicated before it can be passed to the solver as indicator constraints or linearized by the big-$M$ method presented in \cite{rajabdorri2024unit}.

The final step is to modify the objective function and the reserve constraint of \gls{uc} to co-optimize \gls{ufls} alongside operation costs. \Cref{ofmod} presents the modified objective function used here.
\begin{equation}\label{ofmod}
    \min_{u,p}\ c^{\text{g}}(p)+c^{\text{suc}}(u)+c^{\text{\tiny UFLS}}(p^{\text{\tiny UFLS}})
\end{equation}
where,
\begin{equation}\label{uflscost}
    c^{\text{\tiny UFLS}}(p^{\text{\tiny UFLS}})=\sum_t\sum_i c^o_{\text{\tiny UFLS}}\times\text{FOR}_i\times p^{\text{\tiny UFLS}}_{i,t}
\end{equation}
Here, $c^o_{\text{\tiny UFLS}}$ is the post-outage cost of \gls{ufls} in \euro/MW and FOR is the forced outage rate of generators in \%.

Moreover, since an estimation of the \gls{ufls} that will happen after each outage is available now in the optimization problem, the reserve constraint should be modified to schedule less reserve according to the estimation of \gls{ufls}. The reserve constraint used in this formulation is presented in \cref{eq:resanalytical}.

\begin{align}
    \sum\limits_{i\neq \ell}r_{i,t} \geq p_{\ell,t}-p^{\text{\tiny UFLS}}_{\ell,t} &&\text{\footnotesize $\forall t,\; \ell$} \label{eq:resanalytical}
\end{align}

\subsubsection{Data-driven UFLS constraints}\label{mlufls}
An alternative approach to modeling \gls{ufls} inside the \gls{uc} problem and forming a \gls{cfcuc} formulation is data-driven estimation of \gls{ufls} using machine learning techniques. The method implemented in this section is based on \cite{RAJABDORRI2025data}. This method uses the same data set as in \Cref{mlnadir}. The same set of features are selected to represent the data points. However, this time the data points are labeled with the corresponding amount of \gls{ufls} using the \gls{sfr} model. This model can calculate the amount of \gls{ufls} based on the traditional \gls{ufls} schemes used predominately in power systems. This allows the integration of real \gls{ufls} schemes into the problem and a more realistic estimation of \gls{ufls} compared to the analytical formulation.

Using the labeled data set, a regression tree structure is utilized here to estimate the amount of \gls{ufls}. The regression tree employs logistic regression and gives a linear function of features to classify the \gls{ufls} predictions into three different categories. Then, within each resulting leaf, linear regression is used to estimate the values of \gls{ufls}. The tree structure must be as simple as possible, to enable efficient integration into an \gls{milp} formulation.

To model the \gls{ufls} estimation inside the \gls{uc} problem, the tree structure must be represented as \gls{milp} equations. A binary variable is assigned to each leaf allowing activation of one leaf to estimate \gls{ufls}. \Cref{eqleafbin} enforces only one of these binary variables to equal 1. 
\begin{equation} \label{eqleafbin}
    \sum_{\ell e\in\mathcal{L}} z_{\ell e} = 1
\end{equation}

Then the resulting logistic regression linear functions of features after training the model, are used to represent classifications in \gls{milp} form. \Cref{eqclassify} shows the set of constraints utilized for the classification in the first node. Another set of similar constraints must be added for the second classification.
\begin{subequations} \label{eqclassify}
\begin{align}
     \hat{\beta}_0+\sum_{j=1}^J \hat{\beta}_j x_j+\underline{M}\sum_{\ell e\in \mathcal{L}'} z_{\ell e} &\ge \underline{M}\\
     \hat{\beta}_0+\sum_{j=1}^J \hat{\beta}_j x_j+\overline{M}\sum_{\ell e\in \mathcal{L}''} z_{\ell e} & < \overline{M}
\end{align}
\end{subequations}
Where, $\hat{\beta}$ are the logistic regression coefficients, and $\underline{M}$ and $\overline{M}$ are big negative and positive numbers respectively. $\mathcal{L}'$ and $\mathcal{L}''$ are the list of leaves in the right and left subtrees of the node N$_0$.

When the classification is performed, linear regression is used in any chosen leaf to estimate \gls{ufls}. The following constraint calculates \gls{ufls} using linear regression functions.
\begin{equation}\label{eqlinreg}
    p^{\text{\tiny UFLS}}= \sum_{\ell e\in\mathcal{L}}z_{\ell e}\times\left(\alpha_0^{\ell e}+\sum_{j=1}^J \alpha_j^{\ell e} x_j\right)
\end{equation}
Where $\alpha_{j}$ are the linear regression coefficients. \Cref{eqlinreg} must be linearized using the indicator constraint option of the \gls{milp} solvers or the big-$M$ method so that it can be added to the \gls{milp} formulation \cite{rajabdorri2024unit}. 

Like the previous formulation, the objective function and the reserve constraint must be changed. The objective function is changed to \cref{ofmod} to co-optimize \gls{ufls} alongside operation costs. The resulting reserve constraint used in this formulation is presented in \cref{res2}.
\begin{align}\label{res2}
      r_{i,t}\geq \frac{{\hat K}_i u_{i,t}}{\hat K^{s}_{\ell,t}}  (p_{\ell,t}-p^{\text{\tiny UFLS}}_{\ell,t})&&\text{\footnotesize $\forall t,\; \ell$}  
\end{align}
Analogous to the \cref{eqspeed}, \cref{res2} is simplified and then linearized \cite{rajabdorri2024unit}.

\subsection{Definition of models}\label{models}
To clearly distinguish the different models discussed in the previous subsections, each model is clearly defined with the corresponding constraints in \Cref{tab:models}. The maximum \gls{rocof} constraint presented in \cref{rocof} is added to the analytical \gls{cfcuc} formulation to achieve a more fair comparison. It is important to note that the non-linear constraints should be linearized before being added to the formulation, as discussed earlier.
\begin{table}[t!]
    \centering
    \caption{Model definitions}
    \begin{tabular}{M{0.28\linewidth}M{0.2\linewidth}M{0.47\linewidth}}
    \toprule
        model & mutual constraints & unique constraints\\ \midrule
        \gls{buc} &\multirow{5}{*} {\cref{bin1,bin2,ut,dt,pmin,pmax,ru,rd,pb}} & \cref{of,res}\\
        analytical \gls{pfcuc} & & \cref{of,rocof,qss,varc,modnadir}\\
        data-driven \gls{pfcuc} & & \cref{of,rocof,qss,equ_k,eqmlnadir2}\\
        analytical \gls{cfcuc} & & \cref{rocof,eqcrit,eqpufls,ofmod,uflscost,eq:resanalytical}\\
        data-driven \gls{cfcuc} & & \cref{rocof,ofmod,eqleafbin,eqclassify,eqlinreg,res2}\\ \bottomrule
    \end{tabular}
    \label{tab:models}
\end{table}

\section{Comparison framework}\label{benchmark}
Two perspectives are evaluated to compare different \gls{fcuc} thoroughly formulations. First, the system benefits are analyzed in terms of the operation cost, and the total \gls{ufls} for all possible outages. The operation cost and total \gls{ufls} are the main \glspl{kpi} that help compare the effectiveness of the formulations in real operation conditions of island power systems. The amount of \gls{ufls} is calculated using the \gls{sfr} model. The \gls{sfr} model captures the short-term frequency response of a power system using a second-order model approximation to represent the turbine-governor system of the generating units. The excitation and generator transients are neglected due to their faster response times. The detailed model is presented in \cite{sigrist2016island}.

The second perspective analyzed here is the computational burden imposed by each formulation. To compare the computational burden of each model, metrics like the CPU time each formulation takes to reach a solution using a relative
optimality tolerance of 0.1\%, and the compactness of each formulation in
terms of the number of total variables, discrete variables, and constraints are presented.

Additionally, the sensitivity of each model to its relevant threshold is analyzed. The preventive formulations are simulated using different nadir thresholds, and the corrective formulations are simulated with different \gls{ufls} costs. This would show how these formulations can adapt to the level of operation security in various situations.

\section{Results}\label{results}
\subsection{Case study description}\label{casestudy}
All the mentioned formulations are tested using data from a sample day in the power system of a real Spanish island. The generation mix includes eleven diesel generators and about 10\% of its annual demand is provided by wind and solar power generation. Additionally, the real \gls{ufls} settings of the system are used to calculate \gls{ufls} in the \gls{sfr} model so that a realistic comparison of the resulting \gls{ufls} amounts can be performed. Parameters of the diesel generators are provided in \Cref{tab:genparam}. 
\begin{table}[t!]
    \centering
    \caption{Parameters of the Generating Units}
    \begin{tabular}{M{0.07\linewidth}M{0.1\linewidth}M{0.1\linewidth}M{0.1\linewidth}M{0.1\linewidth}M{0.1\linewidth}}
    \toprule
        \textbf{unit} & $P^{\text{max}}$ (MW) & $P^{\text{min}}$ (MW)  & $M^{\text{base}}$ (MVA) & $H$ (s) & $K$ (pu) \\ \midrule
        1 & 3.82 & 2.35 &  5.4 & 1.75 & 20\\
        2 & 3.82 & 2.35 &  5.4 & 1.75 & 20\\
        3 & 3.82 & 2.35 &  5.4 & 1.75 & 20\\
        4 & 4.3 & 2.82 &  6.3 & 1.73 & 20\\
        5 & 6.7 & 3.3 & 9.4 & 2.16 & 20\\
        6 & 6.7 & 3.3 & 9.6 & 1.88 & 20 \\
        7 & 11.2 & 6.63 & 15.75 & 2.1 & 20 \\
        8 & 11.5 & 6.63 & 14.5 & 2.1 & 20\\
        9 & 11.5 & 6.63 & 14.5 & 2.1 & 20\\
        10 & 11.5 & 6.63 &  14.5 & 2.1 & 20\\ 
        11 & 21 & 4.85 &  26.82 & 6.5 & 21.25\\ \bottomrule
    \end{tabular}
    \label{tab:genparam}
\end{table}

The formulations are all simulated in GAMS, and Gurobi 10.0.3 is used as the solver with default settings. The simulations were run on an Intel-i7-8700 CPU @ 3.20GHz with twelve threads and 32 GB of RAM. The code and sample outputs are available on \url{https://github.com/Miadsrv/FrequencyConstrainedUC}.

\subsection{System benefits results} \label{syscompare} 
The simulation results of all the formulations are provided in \Cref{result}. The operation cost of every formulation is given in k\euro \space and the increase/decrease compared to the base case is reported in percentage. Then, the possible generator outages are simulated with the \gls{sfr} dynamic simulator using the real parameters and thresholds of the current \gls{ufls} scheme, and the sum of all possible \gls{ufls} in the sample day is calculated. Similarly, the change in percentage compared to the base case is presented. In addition, the cost of real \gls{ufls} in the corrective formulations is calculated based on the load-shedding costs associated with each scenario. Finally, the average frequency nadir after all possible outages for each formulation is provided.

\begin{table}[t!]
    \centering
    \caption{System benefits results}
    \begin{adjustbox}{width=1\textwidth}
    \begin{tabular}
    {c c c c c c c}\toprule
       case study & $\Delta f^{\text{nadir}}_{\text{crit}}$ (Hz) & $c^o_{\text{\tiny UFLS}}$ (k\euro/MW) & operation cost (k\euro) & $\sum p^{\text{\tiny{UFLS}}}_{\text{\tiny{SFR}}}$ (MW) & $c^{\text{\tiny{UFLS}}}$ (k\euro) & average nadir (Hz)\\ \midrule
       base case \gls{uc} & - & - & 68.15 & 358.25 & - & -1.350\\ \midrule
       \multirow{4}{*}{analytical \gls{pfcuc}} & 3 & - & 71.32 (+4.7\%) & 183.44 (-48.8\%) & - & -1.150\\
       & 2.8 & - & 71.39 (+4.8\%) & 189.63 (-47.1\%) & - & -1.156\\
       & 2.6 & - & 71.46 (+4.9\%) & 181.80 (-49.3\%) & - & -1.159\\
       & 2.5 & - & 72.25 (+6\%) & 178.46 (-50.2\%) & - & -1.171\\ \midrule
       \multirow{4}{*}{data-driven \gls{pfcuc}} & 3 & - & 70.21 (+3\%) & 241.64 (-32.5\%) & - & -1.185\\
       & 2.8 & - & 70.31 (+3.2\%) & 241.58 (-32.6\%) & - & -1.158\\
       & 2.6 & - & 70.81 (+3.9\%) & 202.89 (-43.4\%) & - & -1.141\\
       & 2.5 & - & 70.87 (+4\%) & 200.08 (-44.2\%) & - & -1.151\\ \midrule
       \multirow{4}{*}{analytical \gls{cfcuc}} & - & 0 & 70.11 (+2.9\%) & 219.11 (-38.8\%) & 0 & -1.650\\
       & - & 20 & 70.12 (+2.9\%) & 211.59 (-40.9\%) & 0.97 & -1.158\\
       & - & 50 & 70.20 (+3\%) & 209.27 (-41.6\%) & 2.39 & -1.198\\
       & - & 100 & 70.53 (+3.5\%) & 170.1 (-52.5\%) & 3.88 & -1.294\\ \midrule
       \multirow{4}{*}{data-driven \gls{cfcuc}} & - & 0 & 69.12 (+1.4\%) & 256.93 (-28.3\%) & 0 & -1.198\\
       & - & 20 & 69.55 (+2.1\%) & 161.05 (-55\%) & 0.74 & -1.166\\
       & - & 50 & 69.78 (+2.4\%) & 136.44 (-61.9\%) & 1.56 & -1.078\\
       & - & 100 & 69.91 (+2.6\%) & 130.00 (-63.7\%) & 2.97 & -1.147\\      
      \bottomrule
    \end{tabular}
    \end{adjustbox}
    \label{result}
\end{table}

\begin{figure}[t!]
    \centering
    \includegraphics[width=1\linewidth]{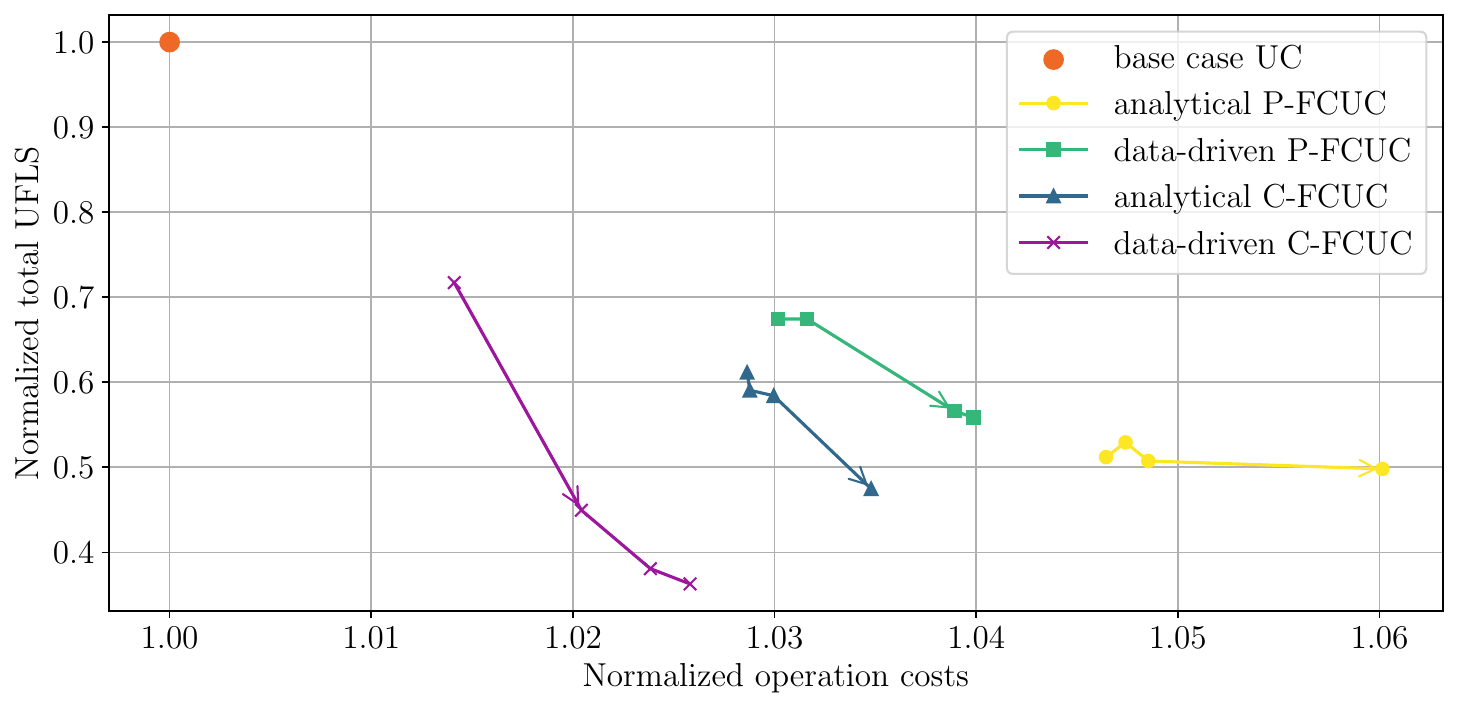}
    \caption{Comparison of system benefits in terms of total \gls{ufls} and operation cost}
    \label{fig:benefits}
\end{figure}

To elaborate on the differences between the formulations, a scatter plot of \glspl{kpi} is presented in  \Cref{fig:benefits}. The numbers are normalized relative to the \gls{buc} to compare the formulations more clearly. For the preventive formulations, the arrow indicates decreasing the nadir threshold, and for corrective formulations, the arrow means increasing \gls{ufls} cost.

Comparing different approaches of preventive formulations, it can be seen that the data-driven approach, gives cheaper operation points but would result in higher \gls{ufls} costs, indicating that the analytical approach is more conservative. An important assumption in the analytical preventive formulation is that all units would provide their reserve linearly in a constant time $T$, which is unrealistic and can cause an underestimation of the frequency drop. 

The comparison between the two different corrective formulations also gives some insights. In both analytical and data-driven approaches it has been ensured that the net power imbalance is distributed among the available units proportional to their primary frequency control speed. It is observed that the data-driven \gls{cfcuc} formulation when assigned with \gls{ufls} cost, has lower cost and lower total \gls{ufls} compared to the analytical \gls{cfcuc}. This is mainly because in the analytical \gls{cfcuc} formulation it is assumed that the \gls{ufls} scheme is adaptive and can be adjusted for each outage. In contrast, traditional \gls{ufls} schemes are employed in practice. This will lead to unrealistic estimations of \gls{ufls} when compared with the resulting \gls{ufls} calculated using the real schemes and parameters of the island power system. In contrast, since the data set used in the data-driven \gls{cfcuc} formulation is trained using the real \gls{ufls} schemes, it has a more realistic estimation of the \gls{ufls}.

Overall, the data-driven \gls{cfcuc} formulation, when assigned with \gls{ufls} cost, performs better in terms of the system benefits compared to the other formulations. The data-driven \gls{cfcuc} has the advantage of having a more realistic estimation of real \gls{ufls} which is key to having lower operation cost and overall \gls{ufls} in real operation.

\begin{figure}[t!]
    \centering
    \begin{subfigure}{0.6\linewidth}
        \includegraphics[width=\linewidth]{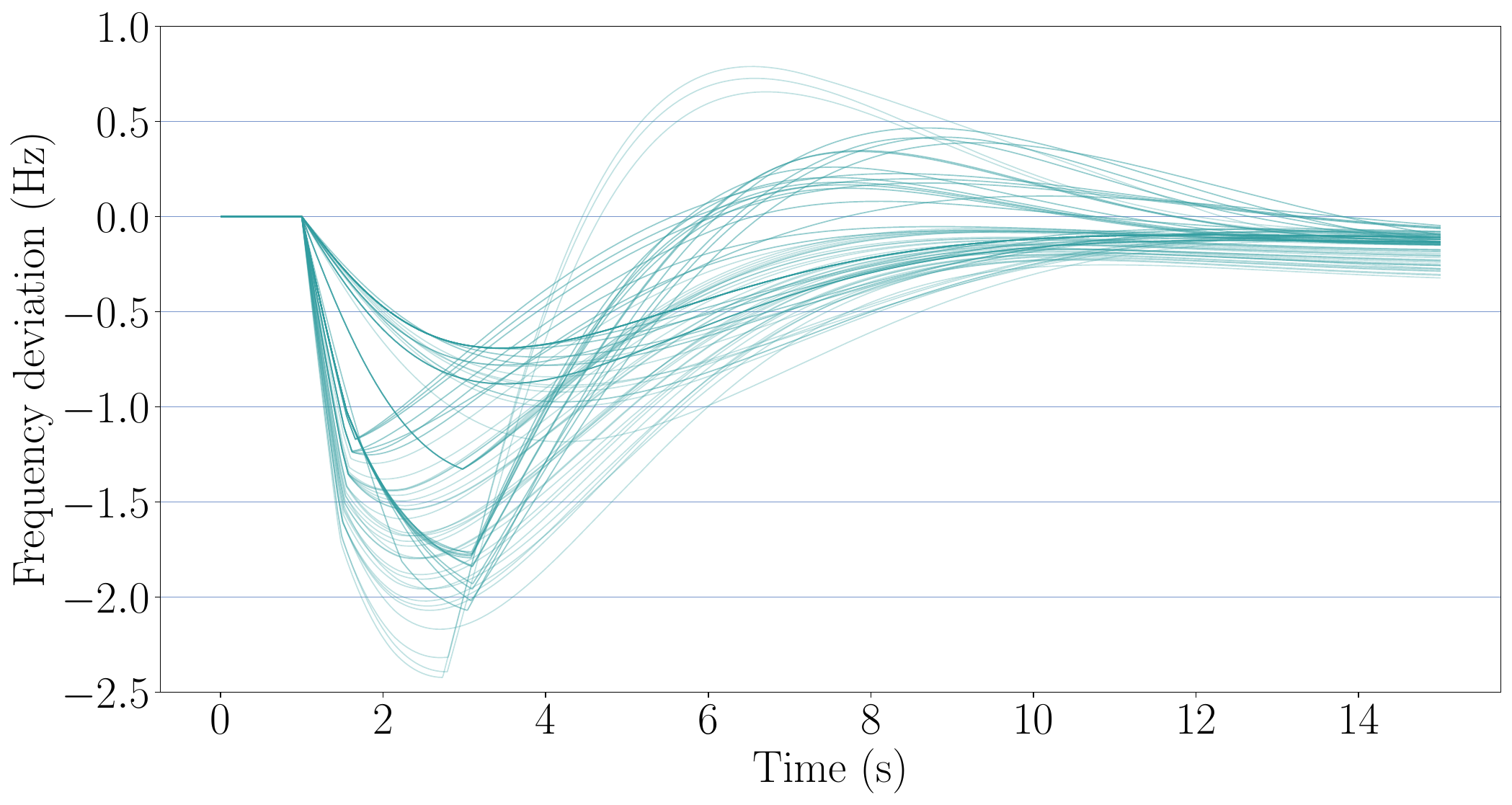}
        \caption{\gls{buc} formulation}
        \label{fig:freqbase} 
    \end{subfigure}
    \begin{subfigure}{0.48\linewidth}
        \includegraphics[width=\linewidth]{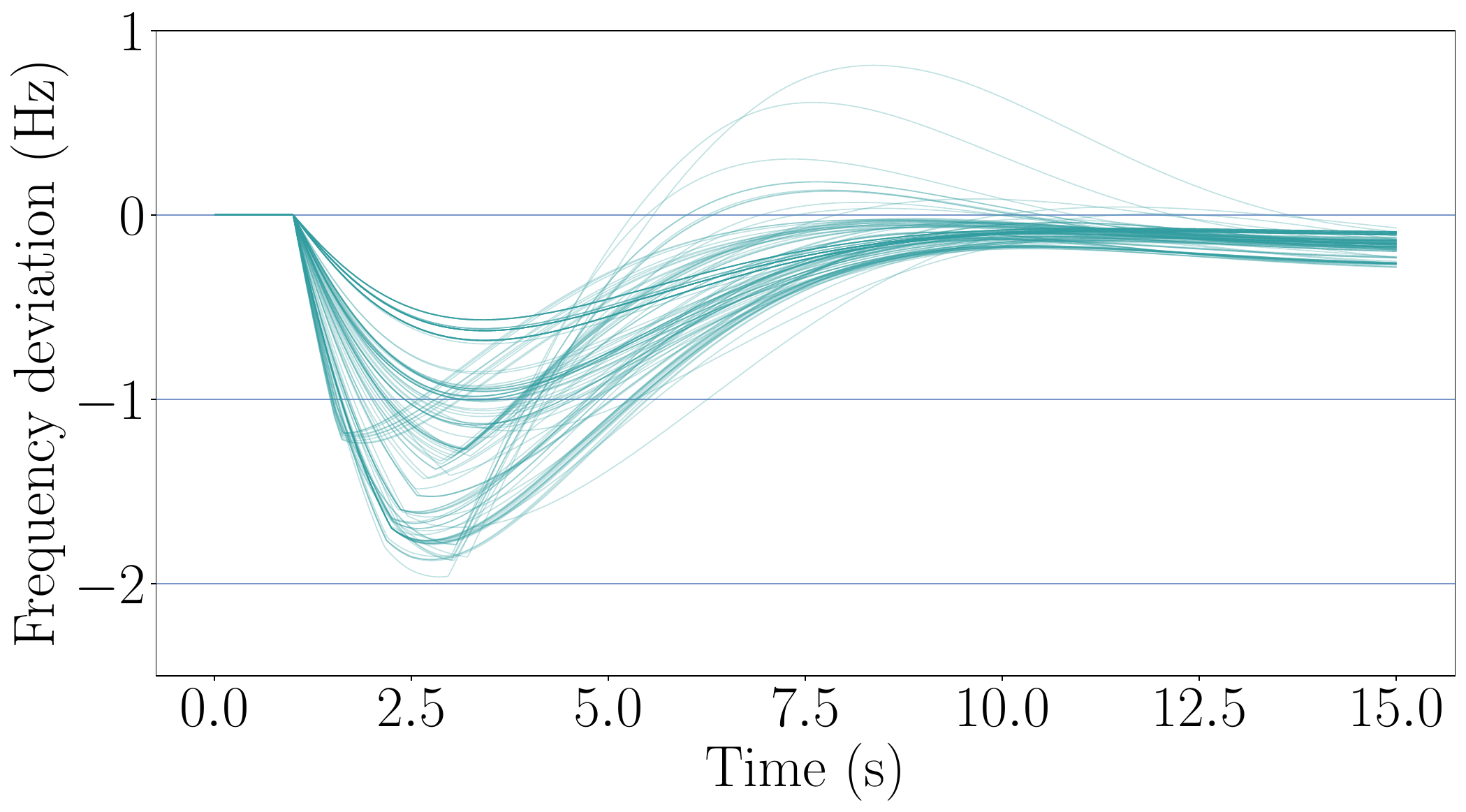}
        \caption{analytical \gls{pfcuc} formulation (2.5 Hz nadir threshold)}
        \label{fig:freqanadir}
    \end{subfigure}
    \begin{subfigure}{0.48\linewidth}
        \includegraphics[width=\linewidth]{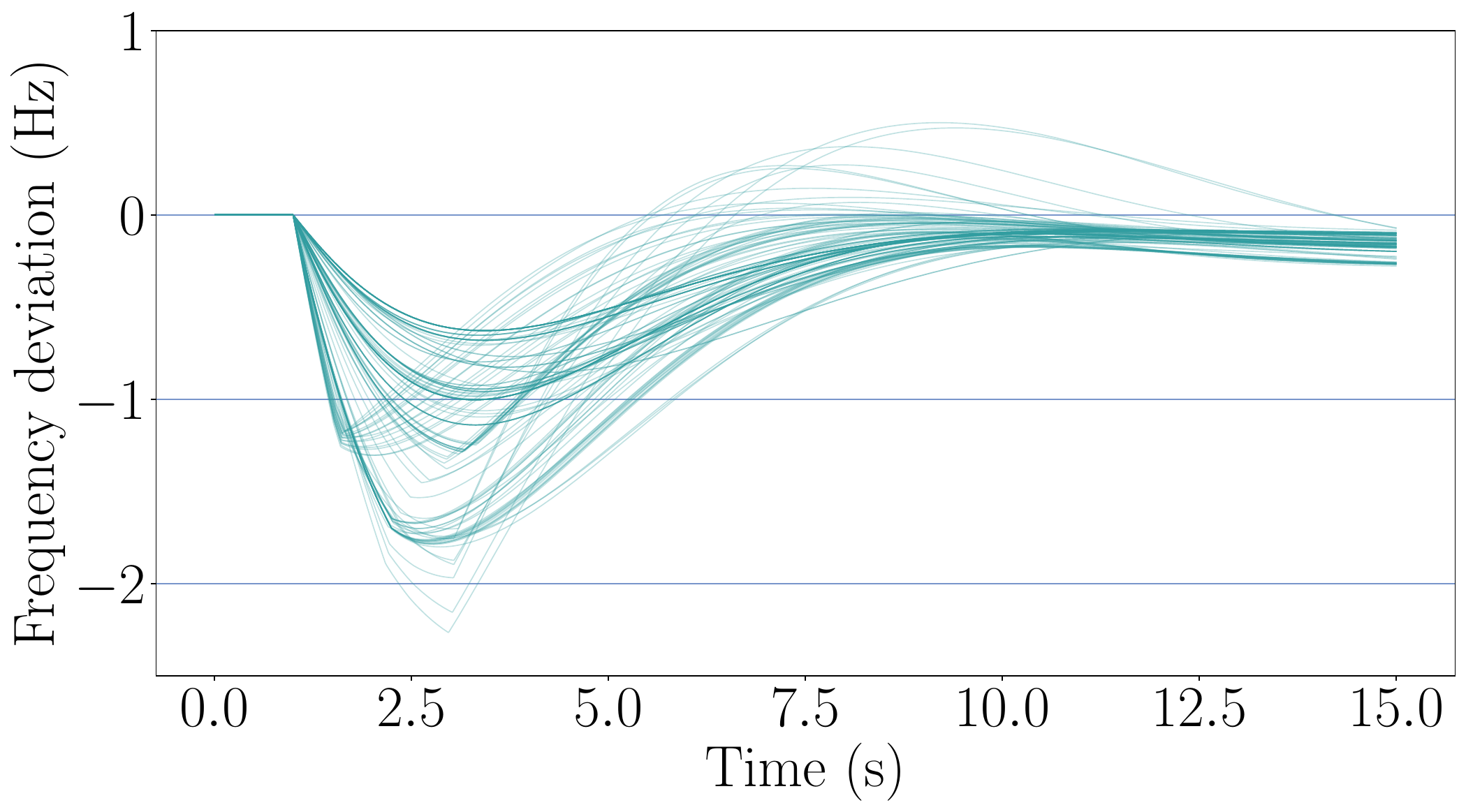}
        \caption{data-driven \gls{pfcuc} formulation (2.5 Hz nadir threshold)}
        \label{fig:freqmnadir}
    \end{subfigure}
    \begin{subfigure}{0.48\linewidth}
        \includegraphics[width=\linewidth]{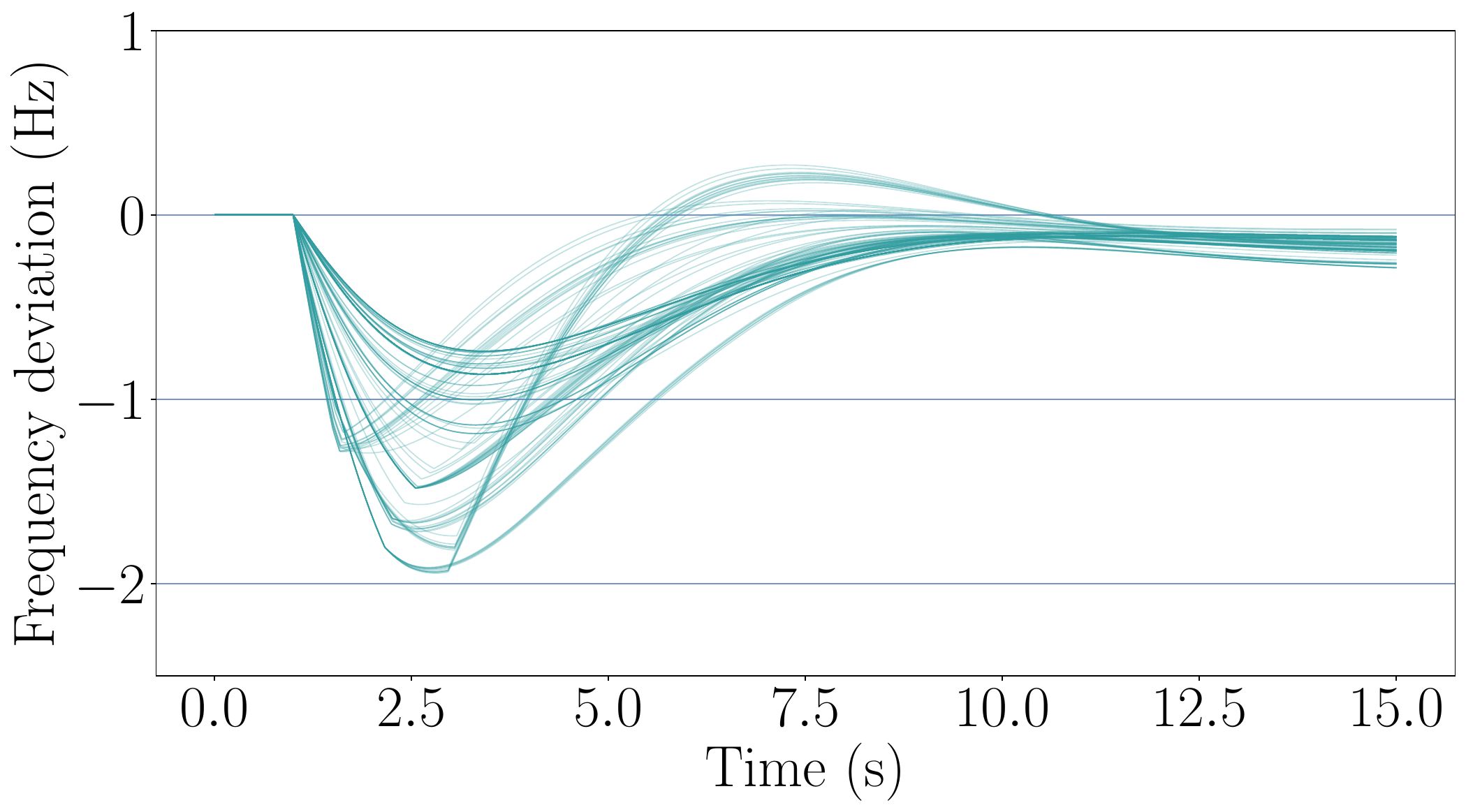}
        \caption{analytical \gls{cfcuc} formulation (50 k\euro/MW cost of \gls{ufls})}
        \label{fig:freqacost}
    \end{subfigure}
    \begin{subfigure}{0.48\linewidth}
        \includegraphics[width=\linewidth]{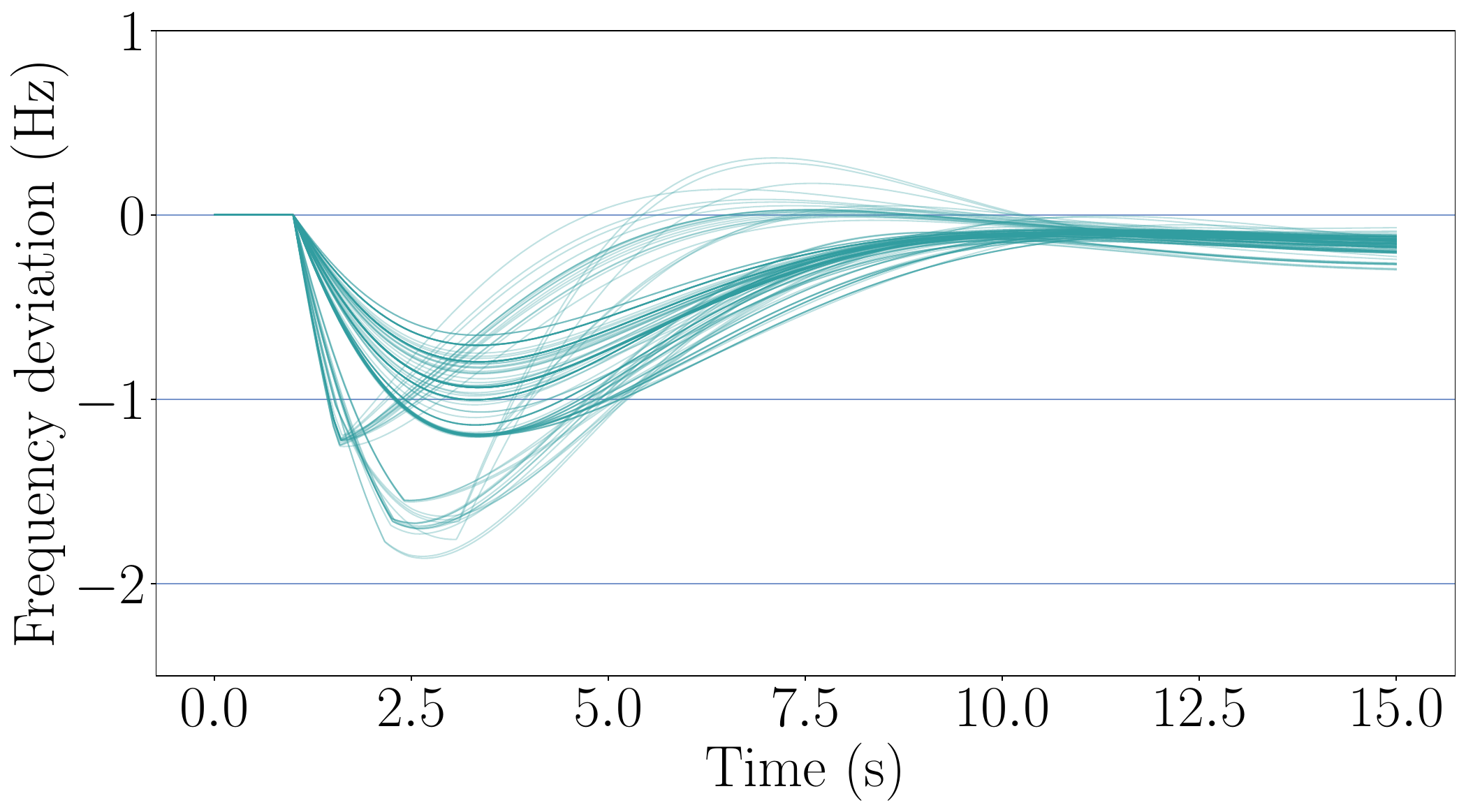}
        \caption{data-driven \gls{cfcuc} formulation (50 k\euro/MW cost of \gls{ufls})}
        \label{fig:freqmcost}
    \end{subfigure}
\caption{Post-contingency frequency performances}\label{fig:freq}
\end{figure}

The resulting frequency responses of dynamic simulations using \gls{sfr} model for each formulation with sample thresholds (2.5 Hz for preventive and 50 k\euro/MW for corrective) and all possible outages are provided in \Cref{fig:freq}.
It is observed in \Cref{fig:freq} that in real operation where conventional \gls{ufls} schemes are utilized, all of the formulations can contain frequency after an outage in a relatively timely manner. According to \Cref{fig:freq}, it is important to note that the \gls{ufls} scheme used in real operation always manages to secure the frequency after outages, and no matter what formulation is used the system stability is not jeopardized.

\subsection{Computational burden results} \label{computcompare} 
\Cref{computational} presents the CPU time, the number of constraints, total variables, and discrete variables for each formulation with different thresholds. It is observed that the corrective formulations generally have higher CPU times as well as more constraints and variables. The preventive formulations perform better in CPU time and compactness of the problem compared to the corrective formulations.

It can be inferred from \Cref{computational} that the analytical \gls{pfcuc} formulation has much higher CPU time compared to the data-driven \gls{pfcuc}. However, the data-driven formulation depends on the data set and takes some time offline to train the model. Also, the data set needs to be updated when the system's structure changes; in that case, the model should be trained again. Changing the nadir threshold in the analytical \gls{pfcuc} formulation from 3 to 2.5 Hz, would result in lower CPU times caused by a tighter solution space. Conversely, this change in the data-driven \gls{pfcuc} formulation would result in higher CPU times.

According to \Cref{computational}, it is noted that the CPU time for the analytical \gls{cfcuc} formulation can range from 8 to 16 minutes for the considered case studies. Additionally, the CPU time in this formulation does not necessarily increase when increasing the \gls{ufls} cost. In contrast, the data-driven \gls{cfcuc} formulation results in CPU times ranging from 2 to 25 minutes, and the solution time increases heavily with increasing the \gls{ufls} costs. Similarly, a disadvantage of the data-driven \gls{cfcuc} is that it depends on the data set, takes some time offline to train the model, and the data set needs to be updated with every change in the system's structure. 

\begin{table}[t!]
    \centering
    \caption{Computational results}
    \begin{adjustbox}{width=1\textwidth}
    \begin{tabular}{c c c c c c c}\toprule
       case study & nadir threshold (Hz) & $c^o$ (k\euro) & CPU time (s) & constraints & total variables & discrete variables \\ \midrule
       base case \gls{uc} & - & - & 7.2 & 6265 & 4513 & 2904\\ \midrule
       \multirow{4}{*}{analytical \gls{pfcuc}} & 3 & - & 167.3 ($\sim$23x) & 15241 & 14281 & 6864\\
       & 2.8 & - & 123.1 ($\sim$17x) & 15241 & 14281 & 6864\\
       & 2.6 & - & 113.1 ($\sim$16x) & 15241 & 14281 & 6864\\
       & 2.5 & - & 104.1 ($\sim$14x) & 15241 & 14281 & 6864\\
       \multirow{4}{*}{data-driven \gls{pfcuc}} & 3 & - & 15.5 ($\sim$2x) & 8113 & 6097 & 2904\\
       & 2.8 & - & 14.3 ($\sim$2x) & 8113 & 6097 & 2904\\
       & 2.6 & - & 18.3 ($\sim$2.5x) & 8113 & 6097 & 2904\\
       & 2.5 & - & 20.1 ($\sim$3x) & 8113 & 6097 & 2904\\ \midrule
       \multirow{3}{*}{analytical \gls{cfcuc}} & - & 0 & 525.7 ($\sim$73x) & 24481 & 14281 & 7128\\
       & - & 20 & 518.8 ($\sim$72x) & 24481 & 14281 & 7128\\
       & - & 50 & 736.1 ($\sim$102x) & 24481 & 14281 & 7128\\
       & - & 100 & 958.8 ($\sim$133x) & 24481 & 14281 & 7128\\
       \multirow{3}{*}{data-driven \gls{cfcuc}} & - & 0 & 146 ($\sim$20x) & 22105 & 11641 & 3696\\
       & - & 20 & 633.9 ($\sim$88x) & 22105 & 11641 & 3696\\
       & - & 50 & 677.1 ($\sim$94x) & 22105 & 11641 & 3696\\
       & - & 100 & 1497.4 ($\sim$208x) & 22105 & 11641 & 3696\\      
      \bottomrule
    \end{tabular}
    \end{adjustbox}
    \label{computational}
\end{table}

\subsection{Overall performance discussion} \label{discuss} 
 
The primary aim is to grasp the formulations' operational and computational aspects simultaneously.
As discussed previously, in the case of small \gls{ips}, the amount of \gls{ufls} is the most important metric that distinguishes the performance of each \gls{fcuc} formulation. Accordingly, two performance indices are introduced. 
The first index presented in \cref{index1} signifies the operational aspect. It is calculated by dividing the increase in operating costs by the total reduction in \gls{ufls} compared to the base case. $c^{\text{op}}$ represents the operation cost corresponding to the formulation. In other words, this index indicates the additional cost incurred in each formulation to reduce one MW of \gls{ufls}.
\begin{equation}\label{index1}
\eta_{s} =
      \frac{c^{\text{op}}-c^{\text{op}}_{\text{\tiny{BUC}}}} {\sum{p^{\text{\tiny{UFLS}}}_{\text{\tiny{BUC}}}}-\sum{p^{\text{\tiny{UFLS}}}}} \\    
\end{equation}

The second index reflecting the computational aspect, is presented in \cref{index2}, where $t^{\text{CPU}}$ is the CPU time of the formulation. Similarly, this index is calculated by dividing the amount that the CPU time increases over the \gls{buc}, divided by the amount of total \gls{ufls} reduction, indicating the amount of time that the solution is prolonged in seconds, to decrease one MW of \gls{ufls}.
\begin{equation}\label{index2}
\eta_{c} =
      \frac{t^{\text{\tiny{CPU}}}-t^{\text{\tiny{CPU}}}_{\text{\tiny{BUC}}}} {\sum{p^{\text{\tiny{UFLS}}}_{\text{\tiny{BUC}}}}-\sum{p^{\text{\tiny{UFLS}}}}} \\    
\end{equation}

\begin{figure}[t!]
    \centering
    \includegraphics[width=1\linewidth]{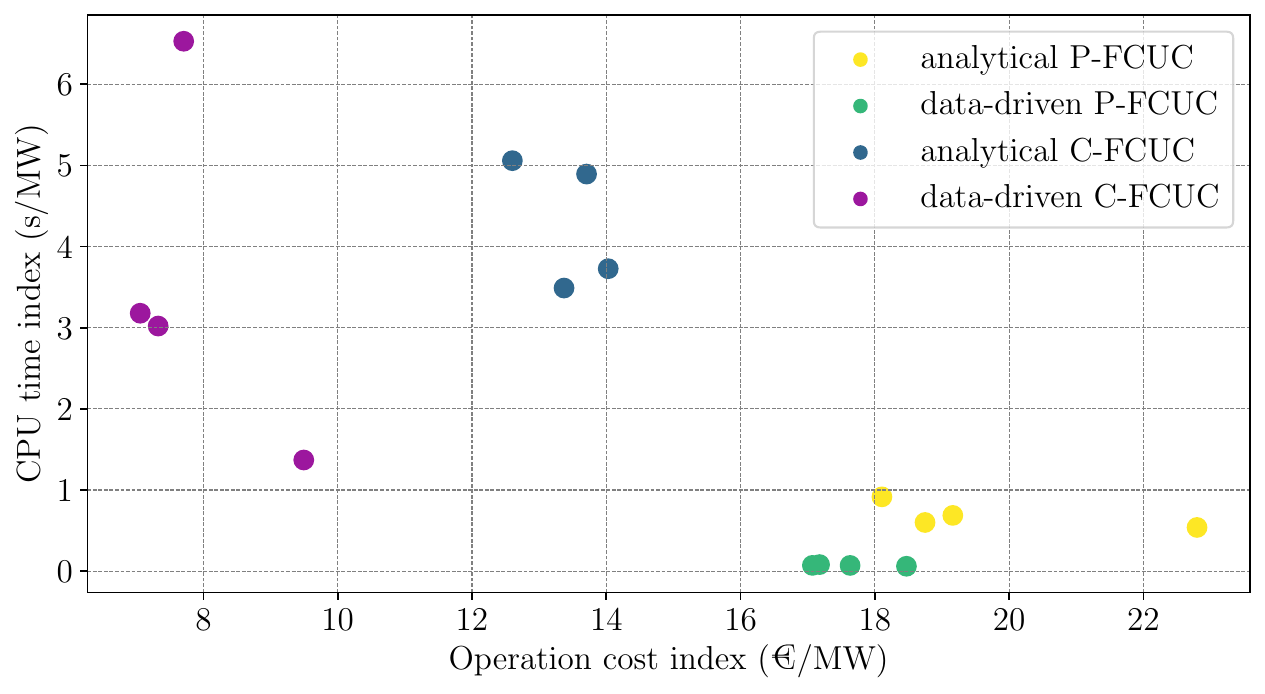}
    \caption{Comparison of the performance indices}
    \label{fig:overall}
\end{figure}

\Cref{indices} gives the numerical results of the performance indices for each formulation. To provide a clearer comparison, \Cref{fig:overall} illustrates the two proposed indices for all formulations. It is noted that the data-driven \gls{cfcuc} formulation shows the best performance regarding the operation cost index, which means that the lowest extra cost should be paid to decrease \gls{ufls}, compared to the other formulations.
\begin{table}[t!]
    \centering
    \caption{Performance indices results}
    \begin{adjustbox}{width=1\textwidth}
    \begin{tabular}{c c c c c c c c}\toprule
       case study & nadir threshold (Hz) & $c^o$ (k\euro) & cost increase (\euro) & CPU time increase (s) & \gls{ufls} reduction (MW) & $\eta_s$ & $\eta_c$ \\ \midrule
       \multirow{4}{*}{analytical \gls{pfcuc}} & 3 & - & 3165 & 160.1 & 174.81 & 18.1 & 0.91\\
       & 2.8 & - & 3231 & 115.9 & 168.62 & 19.2 & 0.69\\
       & 2.6 & - & 3308 & 105.9 & 176.45 & 18.7 & 0.6\\
       & 2.5 & - & 4100 & 96.9 & 179.79 & 22.8 & 0.54\\
       \multirow{4}{*}{data-driven \gls{pfcuc}} & 3 & - & 2056 & 8.3 & 116.61 & 17.6 & 0.07\\
       & 2.8 & - & 2155 & 7.1 & 116.67 & 18.5 & 0.06\\
       & 2.6 & - & 2652 & 11.1 & 155.36 & 17.1 & 0.07\\
       & 2.5 & - & 2717 & 12.9 & 158.18 & 17.2 & 0.08\\ \midrule
       \multirow{3}{*}{analytical \gls{cfcuc}} & - & 0 & 1952 & 518.5 & 139.14 & 14 & 3.73\\
       & - & 20 & 1961 & 511.6 & 146.66 & 13.4 & 3.49\\
       & - & 50 & 2042 & 728.9 & 148.98 & 13.7 & 4.89\\
       & - & 100 & 2371 & 951.6 & 188.15& 12.6 & 5.06\\
       \multirow{3}{*}{data-driven \gls{cfcuc}} & - & 0 & 962 & 138.8 & 101.32 & 9.5 & 1.37\\
       & - & 20 & 1392 & 626.7 & 197.2 & 7.1 & 3.18\\
       & - & 50 & 1625 & 669.9 & 221.81 & 7.3 & 3.02\\
       & - & 100 & 1759 & 1490.2 & 228.25 & 7.7 & 6.53\\      
      \bottomrule
    \end{tabular}
    \end{adjustbox}
    \label{indices}
\end{table}

Alternatively, the data-driven \gls{pfcuc} formulation has the best performance regarding the CPU time index showing that it is the fastest among all the formulations. However, it performs poorly in terms of the operation cost index.

Ultimately, choosing the appropriate formulation depends on the operator's preferences in a given situation. This could involve prioritizing a simpler formulation, minimizing the costs associated with \gls{ufls} if it is costly, or focusing on reducing operating costs.
However, the data-driven \gls{cfcuc} formulation appears to have an advantage in most \glspl{kpi}.
It is important to note that the one data-driven \gls{cfcuc} point corresponding to a high CPU time index is the extreme scenario where \gls{ufls} cost is considered to be 100 \euro/MW, which is an extreme cost and unrealistic \cite{o2021probabilistic}. It is analyzed here only to evaluate the formulation's performance thoroughly. The data-driven \gls{cfcuc} formulation is competent in the CPU time index with other formulations, excluding the mentioned extreme point. Nevertheless, the data-driven \gls{pfcuc} formulation could be utilized when the primary concern is the speed of the formulation.

The advantages and disadvantages of each formulation are summarized in \Cref{tab:procon}.
\begin{table}[t!]
    \small
    \centering
    \caption{Advantages and disadvantages of the evaluated formulations}
    \begin{adjustbox}{width=1\textwidth}
    \begin{tabular}{p{0.18\linewidth}p{0.41\linewidth}p{0.41\linewidth}}
    \toprule
       \centering\textbf{FCUC formulation} & \textbf{Advantages} & \textbf{Disadvantages} \\ 
    \midrule
       \centering Analytical \gls{pfcuc} & 
       \parbox[t]{\linewidth}{\raggedright
       - does not require extra offline training time \newline
       - is faster than corrective formulations \newline
       - results in lower \gls{ufls} than the data-driven \gls{pfcuc}} & 
       \parbox[t]{\linewidth}{\raggedright
       - has the highest operation cost index among all the other formulations\newline
       - some of the assumptions are not realistic.}\\
       
    \midrule
       \centering Data-driven \gls{pfcuc} & 
       \parbox[t]{\linewidth}{\raggedright
       - is the fastest among all formulations and has the best CPU time index \newline
       - is the most compact and adds no binary variable to the \gls{buc} formulation} & 
       \parbox[t]{\linewidth}{\raggedright
       - depends on the training data and needs to be updated according to the structure of the system \newline
       - has an operation cost index higher than the corrective formulations \newline
       - a different model needs to be trained for each threshold} \\
    \midrule
       \centering Analytical \gls{cfcuc} & 
       \parbox[t]{\linewidth}{\raggedright
       - has a better operation cost index than the preventive formulations \newline
       - does not require extra offline training time} & 
       \parbox[t]{\linewidth}{\raggedright
       - has a high CPU time index \newline
       - gives an unrealistic estimation of the \gls{ufls}} \\
    \midrule
       \centering Data-driven \gls{cfcuc} & 
       \parbox[t]{\linewidth}{\raggedright
       - has the best operation cost index among all formulations \newline
       - has a more realistic estimation of the \gls{ufls}} & 
       \parbox[t]{\linewidth}{\raggedright
       - has a higher CPU time index than the preventive formulations \newline
       - depends on the training data and needs to be updated according to the structure of the system} \\     
    \bottomrule
    \end{tabular}
    \end{adjustbox}
    \label{tab:procon}
\end{table}

\section{Conclusions}\label{conclusions}
This paper presents a comparison framework to compare selected formulations of four different \gls{fcuc} methods with a base case \gls{uc} formulation. Such a framework enables system operators of \gls{ips} to choose an appropriate type of formulation to ensure the frequency stability of the system. Additionally, two performance indices were introduced to analyze each formulation's system benefits and computational burden, concerning the single most important criteria in \gls{ips} which is \gls{ufls}. The simulations were conducted on a real Spanish \gls{ips} and each formulation's main advantages and disadvantages are presented.

The results show that regarding the system benefits the data-driven \gls{cfcuc} formulation leads to the lowest operation cost, most \gls{ufls} decrease, and therefore lowest operation cost index. This indicates that the data-driven \gls{cfcuc} formulation can be an ideal formulation when the main goals of the operation are the lowest costs and the lowest total resulting \gls{ufls}.
Moreover, the data-driven \gls{pfcuc} shows the best performance regarding the computational burden and the CPU time index indicating that it could be useful when the main goal is to choose the absolute fastest method among others. However, it will lead to a much higher operation cost index than the corrective formulations.

\section*{Acknowledgment}

This research has been funded by grant PID2022-141765OB-I00 funded by MCIN/AEI/ 10.13039/501100011033 and by “ERDF A way of making Europe”.

\bibliographystyle{elsarticle-num} 
\bibliography{ref}

\end{document}